\documentclass[aps,prl,showpacs,twocolumn,floatfix]{revtex4-1}
\usepackage{amsmath}
\usepackage{amssymb}
\usepackage{amsfonts}
\usepackage{makeidx}
\usepackage{graphicx}
\usepackage{courier}
\usepackage{bm}
\usepackage{color}

\newcommand{\bea}{\begin{eqnarray}}
\newcommand{\eea}{\end{eqnarray}}

\def\be{\begin{equation}}
\def\ee{\end{equation}}
\def\ni{\noindent}

\begin{document} 

\title{Ensemble renormalization group for the random field hierarchical model}

\author{Aur\'elien Decelle$^{1}$, Giorgio Parisi$^{1,2}$ and Jacopo Rocchi $^{1}$}

\affiliation{$^1$Dipartimento di Fisica, Universit\`a La Sapienza, Piazzale Aldo Moro 5, I-00185 Roma, \\
$^2$INFN-Sezione di Roma 1, and CNR-IPCF, UOS di Roma. Italy.}

\begin{abstract}
The Renormalization Group (RG) methods are still far from being completely understood in quenched disordered systems. In order to gain insight into the nature of the phase transition of these systems, it is common to investigate simple models. In this work we study a real-space RG transformation on the Dyson hierarchical lattice with a random field, which led to a reconstruction of the RG flow and to an evaluation of the critical exponents of the model at $T=0$. We show that this method gives very accurate estimations of the critical exponents, by comparing our results with the ones obtained by some of us using an independent method.

\end{abstract}

\maketitle

The Renormalization Group (RG) is a fundamental tool to study the changes of a system as observed at different scales, which has been successfully employed both in quantum field theory \cite{gell1954quantum,*schwinger,*di1969microscopic, *callan1970broken} and in the theory of second order phase transitions \cite{kadanoff1966scaling,*WilsonKG71}.
This process consists in integrating out small scale details of the physical systems: the original interactions between the fundamental degrees of freedom are replaced by \textit{renormalized} interactions between \textit{effective} degrees of freedom. Finally, the critical exponents may be computed repeating this transformation over and over \cite{WilsonKG71, zinn1981perturbation}.
The renormalization group has two main flavours. On the one hand, the process can be done in momentum space, by slowly integrating out high momenta \cite{wilson1974renormalization}. Momentum space RG was first developed in quantum field theory and then it also became a highly developed tool in statistical mechanics, where it is usually performed on a perturbation expansion to compute critical exponents. On the other hand, a technically different approach is to integrate out small distance degrees of freedom in real space \cite{kadanoff1966scaling}. The advantage of the latter is to provide a more physical picture of the process despite the difficulty to obtain accurate results.
In general, both methods need some approximations, but it is possible to find systems with a particular topology for which real space transformations can be performed exactly, such as the two dimensional triangular lattice \cite{niemeijer1976phase} and the diamond hierarchical lattice \cite{BerOst}. These kinds of exactly soluble models could play a very important r\^ole to test new ideas for systems where the nature of the phase transition is difficult to understand.

In this work, we focus on quenched disorder systems for which the renormalization group approach is still not deeply understood. Following the approach introduced with the diamond hierarchical lattice \cite{mckay1982spin, *rosas2001random, *ohzeki2008multicritical}, we concentrate our effort on another hierarchical lattice which gave us the opportunity to define an approximate transformation for more realistic systems.

Years ago, Dyson \cite{dyson1969existence} introduced the so-called Hierarchical Model (HM) to study the problem of phase transitions in one dimensional long-range models. It was later understood that the topology of this model could be used to implement an exact real-space RG transformation \cite{baker1972ising}. 
Therefore, analytical and numerical studies were pursued in this direction \cite{baker1973spin,*gallavotti1975hierarchical,*kim1977critical,*collet1978rg}, \cite{castellana2010hierarchical, *castellana2010renormalization, *castellana2011real}, \cite{decelle2011thesis, angelini2013ensemble,MonthusGarel,ParisiRoc}. Among the numerical works, an approximate real space transformation was suggested as a new approach to deal with the RG in disordered systems. In the first numerical approach \cite{castellana2011real}, a real-space RG transformation for the Spin Glass (SG) model was implemented using as a basis the transformation for the pure model (see also \cite{decelle2011thesis} for more details). However, on the SG problem the estimation of the critical exponent $\nu$ governing the divergence of the correlation length did not always agree with the values obtained by Monte-Carlo (MC) simulation \cite{franz2009overlap}.
Later on, a different RG transformation was proposed in \cite{angelini2013ensemble} for disordered systems leading to values of $\nu$ compatible with MC ones. Still, in order to confirm the validity of this method, large system sizes have to be considered.
Thus, to this aim, we adapt this RG transformation to the random field Ising model on the hierarchical lattice (RFHM), for which we can numerically study large system sizes and compute the critical exponents with high accuracy. Our results can be then compared with the ones obtained in \cite{ParisiRoc}, using an independent method developed in \cite{MonthusGarel}, and they are found to be in good agreement.

The paper is organized as follows. In the first section we define the HM, and discuss some of the main features of the RF models.
In the second section we briefly introduce the RG transformations by defining them for the ferromagnetic model and then generalizing them for the RF model.
We directly illustrate our results by reconstructing the complete phase diagram of the RFHM in the $T/J - h/J$ plane, where $J$ is a coupling constant between spins and $h$ the variance of the random field. 
We show that we do correctly recover the positions of both the transition of the pure model ($h=0,T=T_c)$ and of the random field one ($h=h_c,T=0$), since they agree very well with the results found in \cite{kim1977critical, ParisiRoc}. In the third section we discuss the computation of the critical exponents, comparing them with the values found in \cite{ParisiRoc}.

\noindent \paragraph{The Hierarchical model:} --- The HM is a one-dimensional model where the interaction between spins decreases with the distance defined on a binary tree. The simplest way to define the model is by an iterative construction. First a pair of spins is coupled together with a coupling $0<J_1$. Then a system of four spins is built by coupling two pairs of spins with a new ferromagnetic coupling $0<J_2<J_1$. The operation is then repeated iteratively with these blocks of fours spins using another coupling $0<J_3<J_2<J_1$:

\bea
    \mathcal{H}_1 & = & - J J_1 (s_1+s_2)^2 \label{eq:fsldkfj}\\
	\mathcal{H}_k & = & \mathcal{H}_{k-1}^{(L)} + \mathcal{H}_{k-1}^{( R) } - J \: J_k \left( \sum_{i=1}^N s_i \right)^2 \label{eq_rec_ferro}
\eea

\vspace{-0.7cm}
\begin{figure}[h!]
	\includegraphics[width=6cm,height=1.8cm]{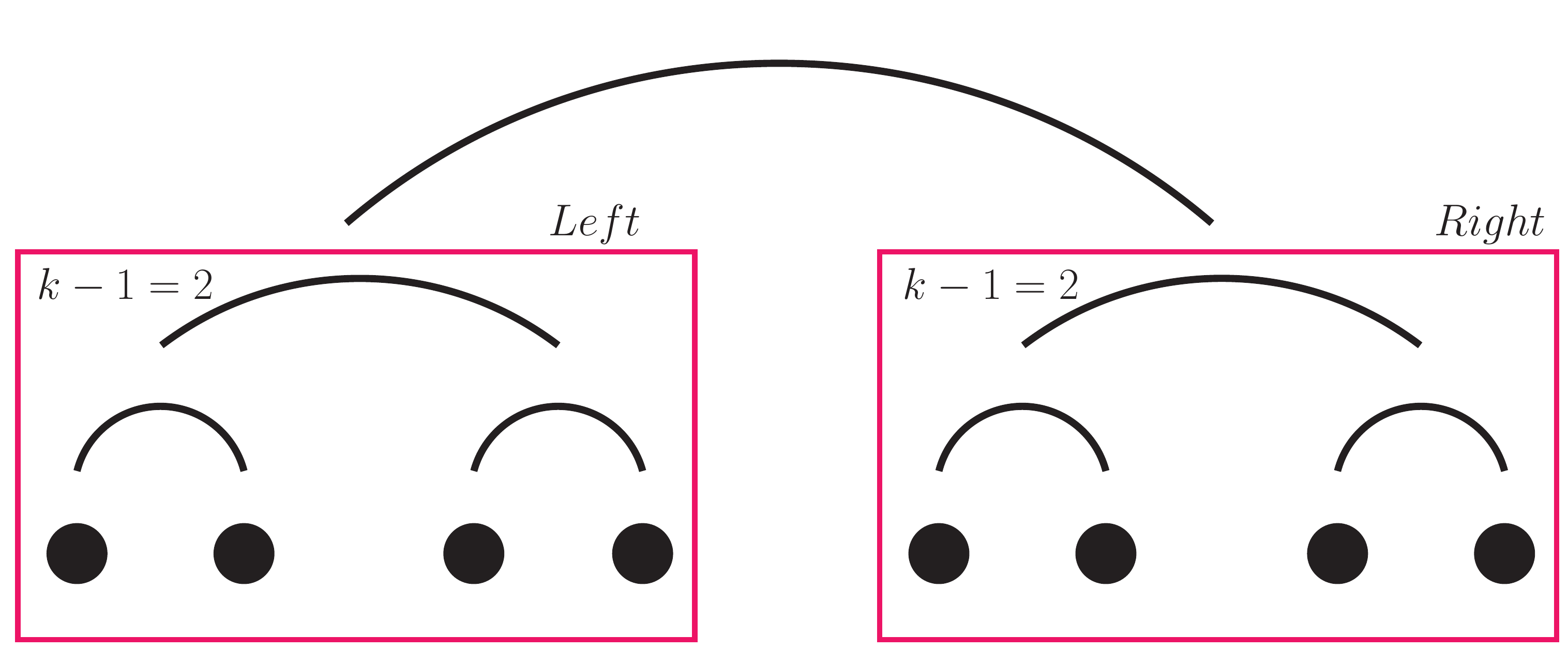}
	\caption{Construction of a system of $n=3$ levels from two smaller ones.}
	\label{fig_HIER}
\end{figure}

\ni where $L$ and $R$ stands for ``left" and ``right" (see Fig. \ref{fig_HIER}). The $J$ and $J_k$ 	are parameters of the model and $N_n \equiv 2^n$ is the system size. In order to approach the behavior of the straightforward one dimensional long range model, where interaction between spins is $J(i,j) = |i-j|^{-\rho}$, it is common to define $J_k=2^{-\rho k}$. The variable $J$ sets the type of interaction (ferromagnetic or anti-ferromagnetic) and its strength. In the following we will always consider $J>0$. The variable $\rho$ controls the strength of the interaction: for $\rho \in [1;2)$ the model has a ferromagnetic phase transition. When $1<\rho<3/2$ the transition is mean field like whereas when $3/2<\rho<1$ it is non-mean field \cite{collet1978rg}. 
Thus, HM can be used to study phase transitions in non-mean field systems, alike to $D-$dimensional short range models, with $D$ smaller than the upper critical dimension (i.e. $D_u^c=4$ for the pure Ising model, $D_u^c=6$ for the RF Ising model and the SG model).
Here the r\^ole of the dimension is played by the continuous parameter $\rho$, which control the decay of the interaction at large distance.
The main advantage of these models is that it is possible to write an exact recursive relation computing the partition function \cite{decelle2011thesis,castellana2011real,angelini2013ensemble, ParisiRoc} in polynomial time with the system size, which can be used to compute correlation functions and other observables.

In this work we consider the random field model on the hierarchical lattice. The RFHM is naturally defined by taking eqs. (\ref{eq:fsldkfj},\ref{eq_rec_ferro})  and adding a random magnetic field of zero mean and variance $h^2$ in the Hamiltonian.
The random field is added only to the first level of the interactions (i.e. in eq. (\ref{eq:fsldkfj})) and therefore relation (\ref{eq_rec_ferro}) still holds. As a consequence, the computation of the partition function and of the correlation functions is still tractable. We can therefore use the same algorithm of the pure model, the only difference being that we have to average over the disorder which increases a bit the complexity. For this model, a simple domain-wall argument suggests that there is has a ferromagnetic phase transition when $\rho \in [1;3/2)$. The situation for $\rho=3/2$ is still not clear. This transition is mean field-like for $\rho<4/3$ and non mean-field  for $\rho>4/3$, \cite{MonthusGarel,ParisiRoc,rodgers1988critical}. As for the pure ferromagnet, similar results hold for the straightforward long range model with random fields, which has been recently investigated in \cite{leuzzi2013imry, hartm2013}. 
In this work, we show that our transformation is suitable to study the random field transition both at $T=0$ and for $T\neq 0$. Indeed the complete phase diagram is characterized by a critical line starting from the pure model and ending in the true critical point at $T=0$. By our method we explain how to recover the full phase diagram of the system in the $T/J - h/J$ plane and how to compute the critical exponent of the true critical point.

\noindent \paragraph{The Ensemble RG transformation:} --- Real space RG transformations can be defined as in \cite{niemeijer1976phase}.
These transformations connect a system $A$ to a decimated system $B$ whose fundamental degrees of freedom can be obtained from the ones of the system $A$ via a coarse-graining procedure.
They are the \textit{effective} degrees of freedom we mentioned in the introduction.
It must be noted that the observables we compute in each system depend on their respective parameters. For simplicity, consider a single parameter $J$.
If we compute an observable in the decimated system, we can use its dependance on $J_B$ and the relation between new spins and the old ones to write an equation $J_B=f(J_A)$.
Suppose that this equation has a fixed point $J^{*}$, then, this equation can be used to compute the critical exponent $\nu$. In fact, as usual in RG theory, $\partial f(J) / \partial J |_{J=J^{*}}=b^{1/\nu}$, where $b$ is the ratio between the size of the system $A$ and the size of the system $B$.
This mapping depends on the parameters of both systems. At the critical point, if the chosen transformation is correct, $J_B=J_A=J^{*}$. This fixed point will be unstable respect to small perturbations from the fixed point value $J^{*}$: if $J_A<J^{*}$, then $J_B<J_A$, while for $J_A>J^{*}$, $J_B>J_A$.

This method works very well in the ferromagnetic hierarchical model, even for small system sizes \cite{castellana2011real, decelle2011thesis}.
In spin glasses, sample-to-sample real-space RG have been tried as in \cite{castellana2011real, decelle2011thesis} but didn't always get satisfying results. Indeed, after obtaining an estimate of the critical exponents, the comparison with Monte-Carlo measures clearly indicate that the method fails in the non-mean field regime.
Recently, Angelini et al. \cite{angelini2013ensemble} proposed a new RG transformation called Ensemble Renormalization Group (ERG). The main difference between this approach and the previous ones came from the order in which the average over disorder and the mapping between the two systems are made.
The general procedure can be described as follows. First we define $\mathcal{N}$ observables $\mathcal{O}_{A,B}^s$ for systems $A$ and $B$, where $s$ runs over different observables and $\mathcal{N}$ is the number of parameters to be determined. Second, we compute $\mathcal{O}_{A}^s$ $\forall s=1,\ldots,\mathcal{N}$. Third, we find the new parameters for the system $B$ such that $\overline{\mathcal{O}^s_{A}}=\overline{\mathcal{O}^s_{B}}$ $\forall s=1,\ldots,\mathcal{N}$.

\begin{figure}
  \includegraphics[width=8cm]{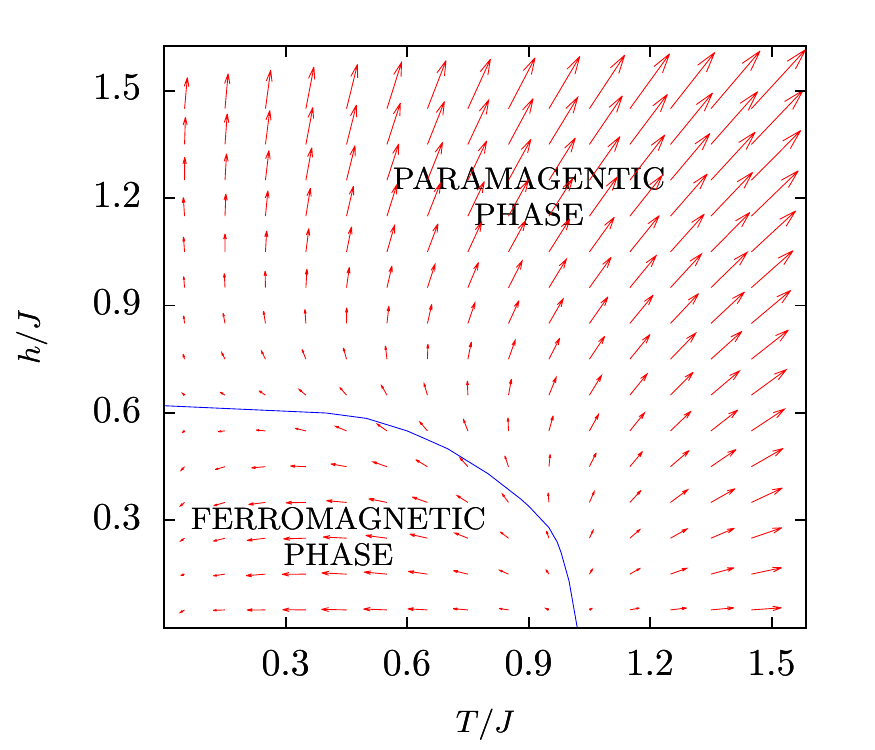} 
  \caption{Phase diagram of the RFHM for $\rho=1.4$ (non-mean field), obtained via the ERG method. The mapping $(J_A, h_A)\rightarrow (J_B, h_B)$ shown in this figure has been obtained with a transformation from systems with $n=5$ levels to systems with $n-1=4$ levels of interactions. For each couple $(J_{A,B},h_{A,B})$ an arrow is drawn. The blue line is a sketch of the critical line inferred by the directions of the vector field.}
  \label{fig_phase_diag}
\end{figure}

We can better describe this method in the hierarchical model.
First let us consider the ferromagnetic case where for each level $k=2,\ldots, n-1$ of the system $A$ (containing $2^n$ spins) we compute 
\be
	O_{k} = \frac{\langle m_{L,k} m_{R,k}\rangle } { \sqrt{ \langle m^2_{L,k} \rangle \langle m^2_{R,k} \rangle} } \:,	
\ee
where $m_{L( R),k}$ is the magnetization of the first (second) block made by $2^k$ spins at the level $k$.
Then, for each level $k=1,\ldots,n-2$ of the system $B$ (containing $2^{n-1}$ spins) we compute the same observables. From the equations $O_{B,k-1}(J_B)=O_{A,k}(J_A)$ for $k=2,\ldots,n-1$ we obtain the mapping $J_B=f(J_A)$.
In principle we just need one observable to determine $J_B$. In practice we observed that the results are much more robust if we enforce that corresponding observables match at each level. In general, this is an impossible requirement since we have more equations than unknowns.
Thus, this condition can be replaced by the weaker one
\be
J_B=\arg \min_{J} \sum_{k=2}^{n-1} \left [O_{B,k-1}(J)-O_{A,k}(J_A) \frac{}{}\right]^2\:.
\label{eq:ergmourea}
\ee
For disordered systems the observables have to be averaged over the many realization of the disorder. 
We have considered here Gaussian distributed random fields for both systems,$ A$ and $B$, parametrized by the variance $h^2$ (the mean being zero).
Equality between corresponding observables is found by fine tuning $h$ and $J$. For the HRFM we consider the two following observables:

\bea
	O^1_{k} &=& \frac{\langle m_{L,k} \rangle \langle m_{R,k} \rangle}{ \sqrt{ \langle m_{L,k}^2 \rangle \langle m_{R,k}^2 \rangle }} \label{eq_tf_1} \:, \\	
	O^2_{k} &=& \frac{\langle m_{L,k} m_{R,k} \rangle - \langle m_{L,k} \rangle \langle m_{R,k} \rangle }  { \sqrt{ \langle m_{L,k}^2 \rangle \langle m_{R,k}^2 \rangle }} \:. \label{eq_tf_2}
\eea

\ni corresponding to both disconnected and connected correlation functions. Similarly to eq. (\ref{eq:ergmourea}), $J_B$ and $h_B$ can be inferred from 
\be
(J_B, h_B)= \arg \min_{J, h}\sum_{s=1}^{2}  \sqrt{ \sum_{k=2}^{n-1} \left[ \frac{}{}d^s_k (J, h; J_A, h_A) \right]^2 }\:,
\label{eq:newJHBBBANBS}
\ee
where we defined
\be
d^s_k (J_B, h_B; J_A, h_A)=\overline{O^s_{B, k-1}}(J_B, h_B) - \overline{O^s_{A, k}} (J_A, h_A) \:.
\ee
This method has been proven to work better than the sample-to-sample RG transformations in the field of SG \cite{angelini2013ensemble}. In fact, sample-to-sample RG transformations tend to reduce the frustration introduced by the disorder, respect to the ERG method, where they are automatically taken into account. Here we exploit this transformation on the RFHM where large system sizes are accessible.

Our method is able to capture the whole phase diagram, as is illustrated on Fig. \ref{fig_phase_diag}. In order to compute the RG flow, we first we pick a point in the ($T/J$, $\sigma/J$) plane, corresponding to the parameters of a system $A$. We then implement a one-step transformation and get a new point ($T/J'$, $\sigma'/J'$) for the system $B$. The temperature acts just as a multiplicative factor. The RG flow is then defined by a vector whose application point corresponds to the initial condition, and whose direction is given by the arrival point (magnitudes have been rescaled to obtain a nicer plot). This procedure may be repeated all over the plane ($T/J$, $\sigma/J$) and allows us to characterize the entire phase space as illustrated for $\rho=1.4$ on Fig. \ref{fig_phase_diag}. Even for modest sizes (i.e. $n=5$) the phase diagram that we obtain is quite good, and the critical line gives the expected values for $h_c$ and $T_c$, apart from finite size corrections \cite{kim1977critical,ParisiRoc} (see \footnote{Respect to the model studied in \cite{MonthusGarel} the couplings $J_k$ have been rescaled with the factor $1-2/2^{\rho}$, thus their critical variance is equal to ours times this factor. This has been done in order to reduce the spread of the region in $h$  where crosses between curves occur.}).


\noindent \paragraph{Estimation of the critical exponent $\nu$:} --- We now describe how to compute critical exponents. In the region $1<\rho<4/3$ the critical exponent $\nu$ takes its mean-field value $\nu=1/(\rho-1)$ \cite{rodgers1988critical}. In another work \cite{ParisiRoc}, numerical estimates for the region $4/3<\rho<3/2$ has been computed using an algorithm developed in \cite{MonthusGarel}. This algorithm provides the ground state configuration of a RFHM sample and can be used to compute the critical exponents by mean of the Finite Size Scaling method.

\begin{figure}
  \includegraphics[width=8cm]{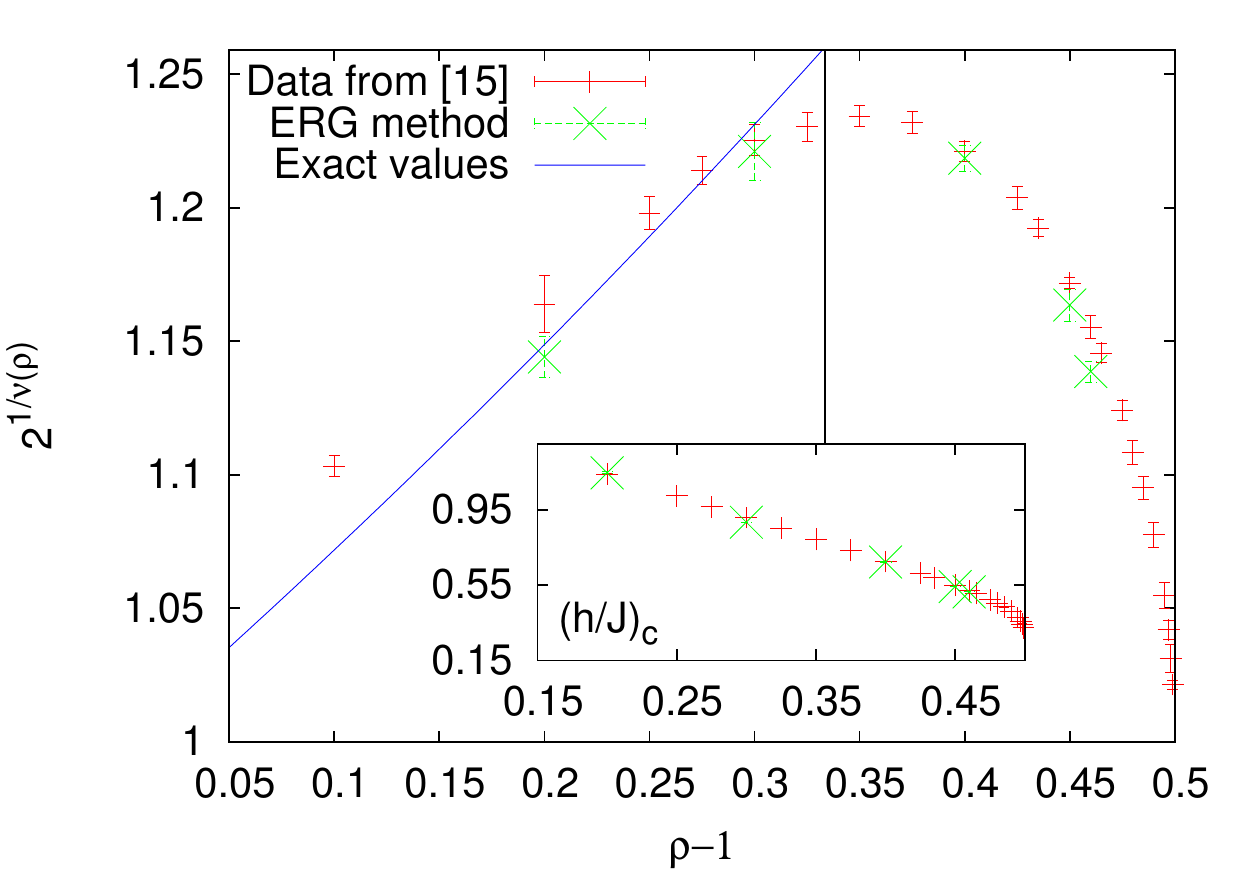}
  \caption{Estimation of $2^{1/\nu}$ using the ERG for different values of $\rho$. We compare these values with the estimations obtained in \cite{ParisiRoc}. We can observe that the two methods agree very well. In the inset, the results for $(h/J)_c$, with error bars (very small). In that case both methods are in complet agreement.}
  \label{fig_nu}
\end{figure}

Our goal here is to both give a quantitative validation of the ERG method by applying it in a disordered system where it is possible to study numerically large system sizes and to estimate the critical exponents of this system in the non-mean field regime. Therefore we finally compare the estimations of the critical exponents given by the ERG method with the values obtained in \cite{ParisiRoc}, see Fig. \ref{fig_nu}. We studied the $T \rightarrow 0$ limit of the ERG transformations \cite{futurwork} and computed the critical exponent $\nu$ and the critical point $(h/J)_c$ for different values of $\rho$. We describe hereafter the details of our method. Given a value of $\rho$, we considered different transformation sizes. Let's consider a size $n$. We first took $M$ samples of such a system and for each, we compute the observables of eq. (\ref{eq_tf_2}). This is the main non--trivial observable in this limit, since the averaged connected correlation function, eq. (\ref{eq_tf_1}), goes to zero as $T\rightarrow 0$.
We then average over the $M$ samples (typically, $M=10^7$ for $n=4$ and $M=10^5$ for $n=11$) and we computed $\overline{O^2}_{A,k}$ for $k=2,\ldots,n-1$. Starting from the initial values $(J_A, h_A)$, we found the values $(J_B, h_B)$ using eq. (\ref{eq:newJHBBBANBS}). This has been done thanks to the C++ open source library \texttt{Dlib} \cite{dlibc}. These transformations have been done for size up to $n=11$.

\begin{figure}
  \includegraphics[width=8cm]{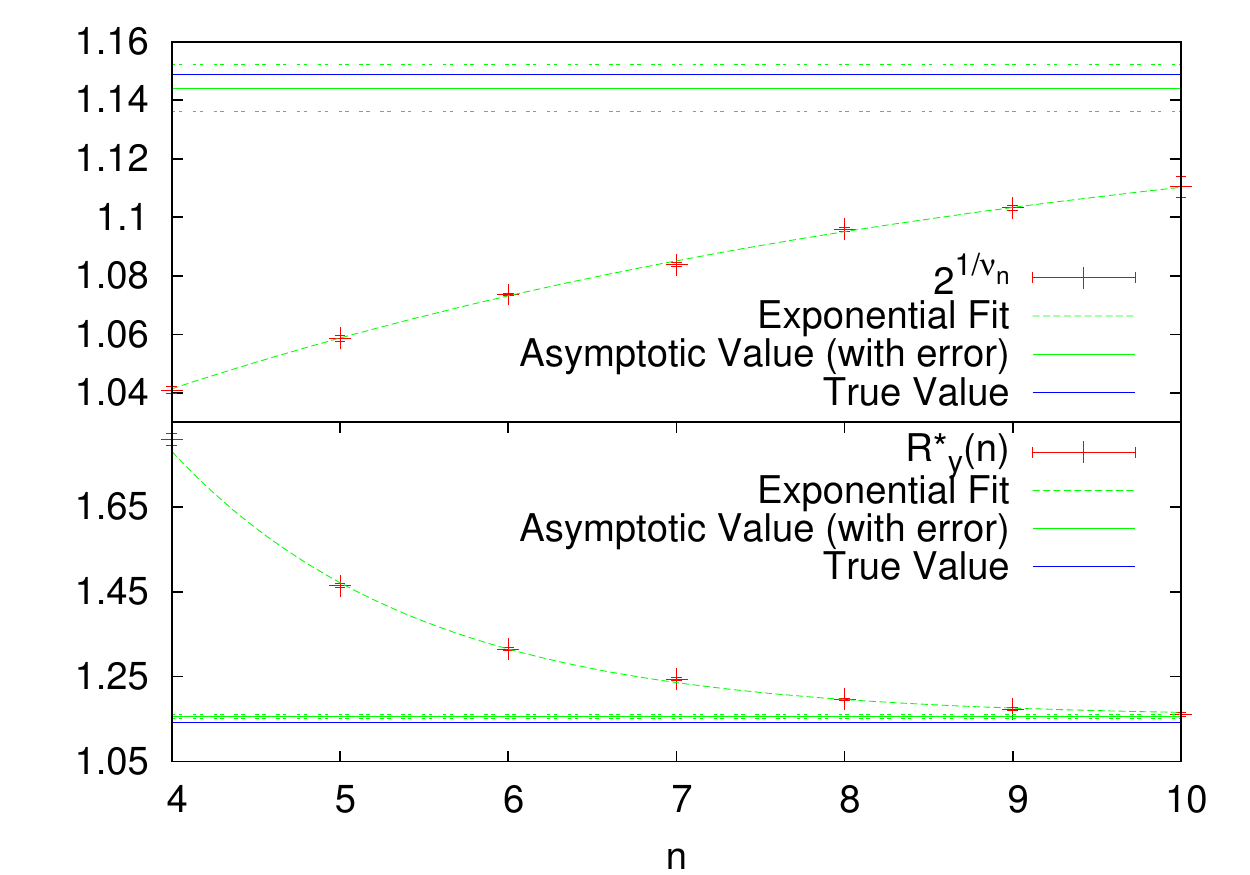}
  \caption{Extrapolation of the value of $2^{1/\nu}$ and $R_y^{c}$ for $\rho=1.2$ (mean-field). These values are obtained using the method described in the text, fitting the data points by an exponential: $f(n)=a+b\: 2^{-c n}$. The extrapolated estimations are very close to the exact parameters of the system, as can be seen from the error bars given in Fig. \ref{fig_nu}. }
  \label{fig_sig}
\end{figure}

In order to measure the dispersion of $(J_B, h_B)$ around their mean values, at a given $(J_A, h_A)$ and $n$, we ran $M'$ times the algorithm (typically $M'=10^2$), each time getting independent estimations of $(J_B, h_B)$. Since we are making the transformation at $T=0$, we only care about the ratio $R_y=h/J$; thus we have to compare $h_A/J_A$ and $h_B/J_B$.
If $R^B_y>R^A_y$ the flow is directed toward the paramagnetic region, while if $R^B_y<R^A_y$, it is directed toward the totally ordered fixed point. 
At the critical point, linearization of the transformation $R^B_y=f(R^A_y)$ allows to measure the critical exponent $\nu$ by using the relation $ \partial R'_y/\partial R_y |_{R_y^{*}} = s^{1/\nu}$ where here $s=2$ since the system B is twice smaller than the system A.
At a given $n$, we obtained curves as the one shown in Fig. \ref{fig_sig}.
%

\ni  The error bars on $\nu$ and $(h/J)_c$ have been obtained by a bootstrap resampling method: the $M'$ pairs of $R_y^{A,B}$ are divided into a finite number of groups and for each group, an estimate of $\nu$ and $(h/J)_c$ is computed. Their dispersions characterize the error bars on $\nu$ and $(h/J)_c$. We studied the ERG transformation $n \rightarrow (n-1)$ for $n=4, \ldots, 11$ and studied finite size corrections like in Fig. \ref{fig_sig}, in order to extrapolate the infinite size limit of the critical ratio $R^c_y$ and the critical exponent $\nu$, see Fig. \ref{fig_nu}.

\paragraph{Conclusion and Acklnowledgements:} --- We developed an RG treatment of the RF problem on the hierarchical topology.
The RG technique we used has been first proposed in \cite{angelini2013ensemble} where they applied it to study spin glasses and dilute magnets. Here, we finally confirm the validity of the method thanks to the possibility to study big systems. In addition, we show that it was possible to recover the whole phase diagram of the model. This study also demonstrates how well the method works, as can be seen on the excellent agreement for the value of the critical point and for the critical exponent.  As a perspective it would be interesting to find a SG model for which the ERG can be implemented for large system sizes. This would also be useful in order to confirm the choice of observables taken in \cite{angelini2013ensemble}. Another development would be to find a possible implementation for other topologies like the Euclidean one.

The research leading to these results has received funding from the European Research Council under the European Unions Seventh Framework Programme (FP7/2007-2013) / ERC grant agreement n. [247328]. A. Decelle has been supported by the FIRB
project n. RBFR086NN1. We would also like to acknowledge stimulating discussions with F. Ricci-Tersenghi, David Yllanes and Beatriz Seoane.

\bibliography{Bib}
\end{document}